# Energy use in quantum data centers: Scaling the impact of computer architecture, qubit performance, size, and thermal parameters

Michael James Martin, Caroline Hughes, Gilberto Moreno, Eric B. Jones, David Sickinger, Sreekant Narumanchi, and Ray Grout

National Renewable Energy Laboratory, Golden, CO, 80401.

E-mail: Michael.Martin@nrel.gov.

As quantum computers increase in size, the total energy used by a quantum data center, including the cooling, will become a greater concern. The cooling requirements of quantum computers, which must operate at temperatures near absolute zero, are determined by computing system parameters, including the number and type of physical qubits, the operating temperature, the packaging efficiency of the system, and the split between circuits operating at cryogenic temperatures and those operating at room temperature. When combined with thermal system parameters such as cooling efficiency and cryostat heat transfer, the total energy use can be determined. Using a first-principles energy model, this paper reports the impact of computer architecture and thermal parameters on the overall energy requirements. The results also show that power use and quantum volume can be analytically correlated. Approaches are identified for minimizing energy use in integrated quantum systems relative to computational power. The results show that the energy required for cooling is significantly larger than that required for computation, a reversal from energy usage patterns seen in conventional computing. Designing a sustainable quantum computer will require both efficient cooling and system design that minimizes cooling requirements.

**Index Terms**—Cryogenics, data center integration, energy efficiency, quantum computing, sustainability

## 1. Introduction

Quantum computers capable of solving the scientific and cryptography problems that drive research interest in the technology will be much larger than current systems. For instance, a computer capable of using Shor's algorithm for current cryptography problems will require on the order of $10^6$ physical (as opposed to computational) qubits,[1] while using quantum computing to simulate nitrogen fixation, a computational chemistry problem widely used as an example of the scientific potential of quantum computing, will require on the order of $10^5$ to $10^8$ physical qubits, depending on the error rate.[2] This compares to the 53-qubit machine recently used to sample the output of a pseudorandom quantum circuit.[3] As computers scale up, their power usage will also increase.

Because of the extremely low temperatures required by quantum computing, the power consumption of the cooling system is likely to be much greater than the power consumption of the electronics. This has already been demonstrated with existing quantum annealing systems,[4] and was identified as a major challenge in a recent National Academies of Sciences, Engineering, and Medicine overview of the state of quantum computing.[5] These challenges are not limited to quantum computing; they have been identified as a limiting factor for incorporating other cryogenic components into conventional computing architectures.[6-7] These changes in the balance of power supplied to electronics compared to the power for cooling mark a significant change from conventional computing, where cooling requirements are typically 10%–30% of the power used by the electronics.[8] Identifying the combined cooling and electronics power requirements of these systems is therefore crucial to both engineering quantum computers to decrease power consumption, and predicting the impact of quantum systems on national and global energy use.[9-10]

## 2. Energy Model Development

### 2.1. Model Derivation

As shown in Fig. 1, quantum information systems are *heterogeneous* computer systems[11] composed of (1) qubits and associated circuits and sensors operating at a temperature $T_c$ in a cryostat, and (2) external control and communications circuits operating at a temperature $T_H$ at or above the ambient temperature $T_o$. There are two electronic power loads: $P_1$ is the power used by the qubits and other circuits operating at cryogenic temperatures, and $P_2$ is the power used by the external circuits. These terms can be combined to form a total electronic power load $P_c$.

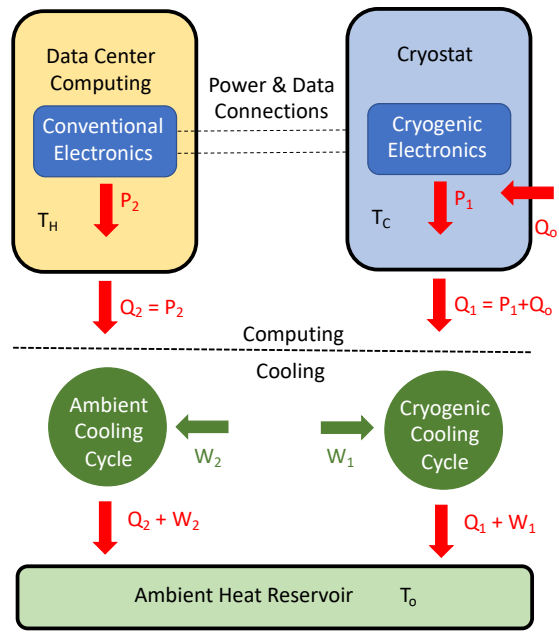

Fig. 1: Quantum Computing Data Center Model and Energy Movement

Scaling these power loads based on the size of the system is complicated by the range of electronics that will operate in the system. In addition to the qubits and their associated control and readout electronics, most quantum systems will contain a set of gates and other associated electronics. Their power use must also be taken into account.

A reasonable first approximation is that the total power load will scale as some power $a$ of the number of physical qubits in the computer $n_p$:

$$P_c \propto (n_p)^a. \tag{1}$$

The most likely scaling is a linear scaling, where $a$ is equal to one. In this case $P_c$ will be equal to $n_p q$, where $q$ is the value of energy used per qubit:

$$P_c = n_p q \tag{2}$$

The power use can then, at both the qubit and overall system level, be split between the power used in the cryostat and the ambient environment. The power used per qubit $q$ will be split between the power used per qubit by the electronics inside the cryostat $q_1$, and the power used per qubit by the electronics at ambient temperature $q_2$:

$$q = q_1 + q_2. \tag{3}$$

The parameter $\phi$ can then be used define how the total energy is split between the circuits operating at cryogenic temperatures and the external circuits:

$$q_1 = \phi q, \tag{4}$$

$$q_2 = (1 - \phi)q, \tag{5}$$

$$P_1 = \phi P_c, \tag{6}$$

$$P_2 = (1 - \phi)P_c. \tag{7}$$

A value of $\phi$ of 1 then corresponds to a system where all electronics operate at cryogenic temperatures, while a value of 0 represents a system where there is no energy usage in quantum systems.

To maintain the cryogenic electronics at a temperature $T_c$ that is below the ambient temperature $T_o$, the cryogenic cooling system must remove both the power dissipated by the circuits in the cryostat, equal to $P_1$, and the heat entering the chamber $Q_o$. $Q_o$ will depend on the cryogenic chamber area $A$, the heat transfer coefficient $U$, and the temperatures $T_c$ and $T_o$:

$$Q_o = UA(T_o - T_c). \tag{8}$$

Because $T_c$ is much less than $T_o$, $Q_o$ can be approximated as being equal to $UAT_o$. Two additional assumptions allow this value to be rewritten in terms of the number of qubits. The first assumption is that $A$ is equal to $C_l V^{2/3}$, where $V$ is the volume of the chamber and $C_l$ is a geometric constant. (For a cube, $C_l$ is 6; for a sphere, $C_l$ is $\sqrt[3]{36\pi}$.) The second assumption is that volume can be re-written as the product of the number of qubits and a volume per qubit $v_q$, so that $V$ is equal to $n_p v_q$. Note that $v_q$ is not the volume of the qubit, but the volume required to house a qubit when all interconnections and other geometric constraints are taken into account. The heat load $Q_o$ now depends on the number of qubits:

$$Q_o = UC_1 T_o v_q^{2/3} n_p^{2/3}. \tag{9}$$

The total heat $Q_1$ that must be removed from the cryogenic chamber $Q_c$ is the sum of $P_1$ and $Q_o$:

$$\begin{aligned} Q_1 &= P_1 + Q_o = n_p q_1 + UC_1 T_o v_q^{2/3} n_p^{2/3} \\ &= n_p q_1 \left(1 + (UC_1 T_o v_q^{2/3}/q_1) n_p^{-1/3}\right) \\ &= n_p q_1 \left(1 + \beta n_p^{-1/3}\right) \\ &= \phi n_p q \left(1 + \beta n_p^{-1/3}\right), \end{aligned} \tag{10}$$

where the parameter $\beta$ equal to $UC_1 T_o v_q^{2/3}/q_1$ captures the importance of heat transfer into the chamber relative to the electronic heat generation within the chamber. This scaling has one limitation: it assumes that power of the electronics in the cryostat is not negligible compared to the heat entering the cryostat, which would lead to a value of $\beta$ of infinity. This situation could occur if the packaging is not optimized to minimize heat entering the cryostat. Conversely, a perfectly insulated system would have a $\beta$ of zero.

The work $W_1$ required to remove heat from the cryogenic chamber is calculated by dividing the heat by the coefficient of performance, or $COP(T_C)$.[12] This value can be expressed as a product of the value of the Carnot efficiency $COP(T_C)|_C$ and a correction factor $\eta_c$:

$$COP(T_c) = \frac{Q_1}{W_1} = \eta_c COP(T_c)|_C = \eta_c \left(\frac{T_c}{T_o - T_c}\right). \tag{11}$$

Combining (10) and (11) gives a value for $W_1$:

$$W_1 = n_p q_1 \left(\frac{1 + \beta n_p^{-1/3}}{\eta_c COP(T_c)|_C}\right) = \phi n_p q \left(\frac{1 + \beta n_p^{-1/3}}{\eta_c COP(T_c)|_C}\right). \tag{12}$$

The heat that must be removed from the external electronics $Q_2$ is equal to $P_2$. Because the cooling of room-temperature electronics is not performed using a direct refrigeration cycle, the cooling energy cost $W_2$ is characterized using a cooling figure of merit, which is defined similarly to a coefficient of performance:

$$W_2 = \frac{P_2}{FOM(T_o)} = \frac{n_p q_2}{FOM(T_o)} = \frac{(1-\phi) n_p q}{FOM(T_o)}. \tag{13}$$

The total energy required to cool the system $W_S$ is then the sum of $W_1$ and $W_2$:

$$\begin{aligned} W_S &= n_p q_1 \left(\frac{1 + \beta n_p^{-1/3}}{\eta_c COP(T_c)|_C}\right) + \frac{n_p q_2}{FOM(T_o)} \\ &= n_p q \left[\phi \left(\frac{1 + \beta n_p^{-1/3}}{\eta_c COP(T_c)|_C}\right) + \left(\frac{1-\phi}{FOM(T_o)}\right)\right]. \end{aligned} \tag{14}$$

The total power required by the system $P_T$ is then the sum of $P_C$ and $W_S$:

$$P_T = n_p q \left[1 + \phi \left(\frac{1 + \beta n_p^{-1/3}}{\eta_c COP(T_c)|_C}\right) + \left(\frac{1-\phi}{FOM(T_o)}\right)\right]. \tag{15}$$

## 2.2. Derivation of scaled energy usage

Because $q$ is the most difficult of these terms to estimate for future quantum information systems, it is useful to look at the ratio of the energy use $P_T$ to the total computational power usage $q$ required for a single qubit, which is defined as $P_T^*$:

$$P_T^* = n_p \left[ 1 + \phi \left( \frac{1 + \beta n_p^{-1/3}}{\eta_c COP(T_c)|_C} \right) + \left( \frac{1 - \phi}{FOM(T_o)} \right) \right]. \tag{16}$$

Note that $q$ is not completely eliminated from the equation: it still appears in the scaling of $\beta$. However, if $\beta$ is seen fundamentally as a scaling of the importance of heat dissipation within the cryostat to heat entering the cryostat, this simplification remains useful and appropriate. The identical scaling can be applied to $P_1$, $P_2$, $Q_c$, $W_2$, $W_3$, $P_S$, and $P_c$.

A range of metrics are available for quantifying the energy use and sustainability of data centers.[13] The efficiency of conventional data centers is frequently characterized by the power usage efficiency, or PUE.[14] This metric is the ratio of the total energy used by the data center $P_T$ divided by the power going to computation $P_C$. If other possible loadings such as lighting are excluded, the PUE for a quantum system can be found simply by dividing (15) by a value of $P_C$ of $n_p q$:

$$PUE = 1 + \phi \left( \frac{1 + \beta n_p^{-1/3}}{\eta_c COP(T_c)|_C} \right) + \left( \frac{1 - \phi}{FOM(T_o)} \right). \tag{17}$$

## 3. Review of Representative Physical Parameters

Though there is considerable uncertainty in how quantum computers will evolve, enough is known about key parameters to allow initial assessments of energy use in quantum computing. As shown in Fig. 2, several key parameters can already be identified for quantum systems. Two types of qubits have already been integrated into quantum computers: superconducting qubits and ion-trap qubits; each has a range of reported operating temperatures.[15-20] The temperature ranges for silicon and diamond qubits, which have been demonstrated as stand-alone components, are also shown.[21-25] The total operating range for existing qubit technologies ranges from 0.01 K to 10 K. The type of qubit used will determine $T_c$ and $COP_c(T_c)$ for the system.

At 10 mK, distillation refrigeration using helium and laser cooling are currently the only viable cooling technologies. Laser cooling has been successfully used to maintain ion-trap qubits at 4 K.[18] However, in general, laser cooling systems have low thermodynamic efficiencies: though theoretically up to 20% of the Carnot efficiency is achievable, in practice only 3% has been achieved.[26] Distillation refrigerators have been used on a range of large- and small-scale scientific instrumentation to produce low-temperature helium II. Figure 2 shows the values of COP for both large-scale[27-28] and chip-scale[29-30] cooling systems. While chip-scale systems show a large drop in efficiency as the temperature decreases, the efficiency of large-scale systems appears to remain between 12% and 18% of the Carnot efficiency, even when the system temperature is below the boundary between helium I and helium II.

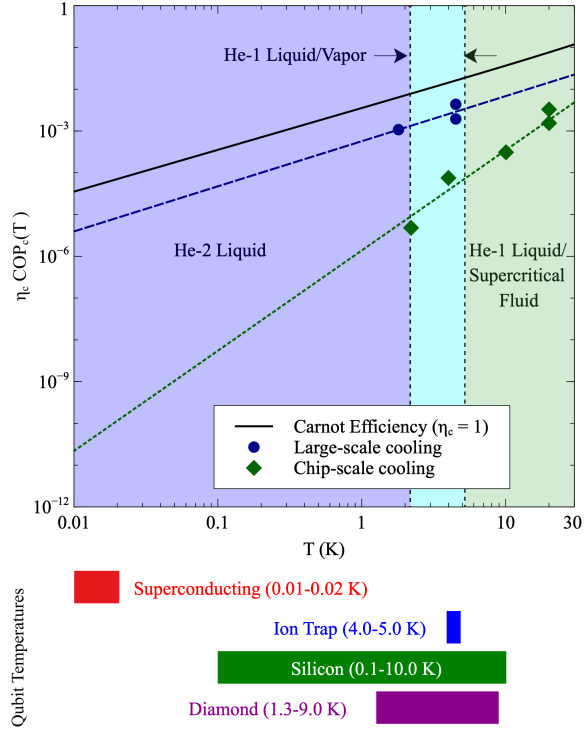

Fig. 2. Quantum data center operating temperatures and cryogenic cooling efficiencies with relevant helium temperature ranges.

For purposes of this study, $\eta_c$ is estimated to be 0.15 for the entire range of 10 mK–10.0 K. The value for the *FOM*($T_o$) can be estimated based on experience with conventional data centers. An efficiently designed data center will typically have a PUE of 1.2, which corresponds to a value for *FOM*($T_o$) of 5.0, which is used in this study. The remaining thermal parameter, $U$, is estimated to be 0.3–0.8 mW/m$^2$-K, based on experience with existing cryostats used in the Large Hadron Collider.[31]

The only computer architecture parameter that can be identified that would be relevant for future systems is the number of physical qubits $n_p$ to perform useful computational tasks. The required power for electronics per qubit $q$ that will be seen in future quantum systems is challenging to estimate. The Landauer limit suggests the value of modifying information will be proportional to, or will at least increase with, $T_c$.[32] Though this limit has been extended to quantum systems,[33] it does not reflect the power usage of related electronics either inside or outside the cryostat. Additional work suggests a lower bound for energy use per qubit,[32] while other research suggests that proper manipulation of information in quantum systems can lower the total energy used.[35] However, the value of $q$ for future quantum systems cannot be reliably determined. Similarly, the required cryostat volume per qubit, $v_q$, cannot be characterized for a rapidly changing technology.

Though $\phi$ is bounded between 0 and 1, it may vary by orders of magnitude within that range in future systems. The value of $\phi$ reflects not only the qubit type and the power required by associated sensors, control systems, and electronics, but also a design decision about where to place conventional silicon electronics used to control the qubits. Placing these electronics inside the cryostat may simplify integration and decrease latency,[36] but doing so leads to reliability challenges for CMOS electronics.[37] Doing so also creates a cooling penalty that has already been observed in laboratory experiments.[38] These reliability and power concerns have led to the development of specialized conventional CMOS circuits for use with quantum electronics.[39] The design decisions made to move electronics in or out of the cryostat, increasing or decreasing $\phi$, will also increase or decrease $v_q$, as more cryostat volume and surface heat transfer area will be required per qubit as these electronics are added.

### 4. Energy Use Results

In spite of the inability to assign values to many parameters for quantum systems, the power used for integrated quantum systems can still be studied using values of $P^*$ and $PUE$, in which $q$ and $n_p q$ respectively are factored out. The impact of different values of $\phi$ and of $\beta$, as well as system operating temperatures and size, can then be explored.

### 4.1. Ideal Results: No heat transfer

The limiting case for maximum system efficiency will be a perfectly insulated cryostat, which corresponds to a value of $\beta$ of zero. In this case, $P^*$ is proportional to $n_p$, and the PUE is independent of the size of the computer. Figure 3 shows the PUE as a function of $\phi$ for the two types of qubits that have been successfully integrated into quantum computers for the range of operating temperatures $T_c$ demonstrated for both qubit types.

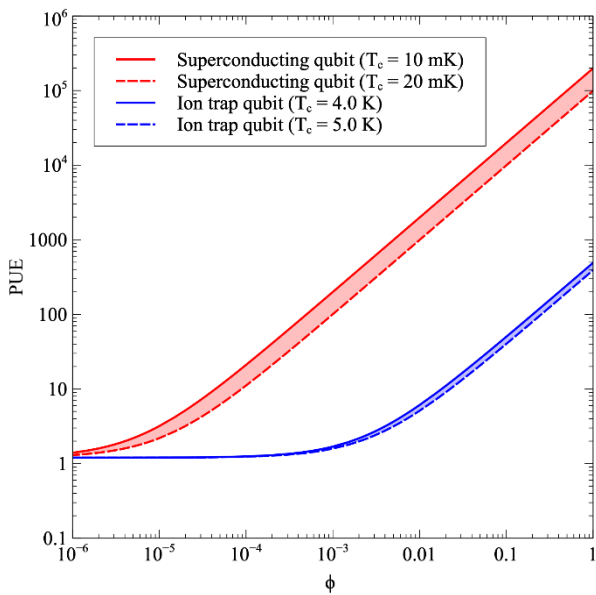

Fig. 3: Quantum Data Center Power Usage Efficiencies (PUEs) with no external heat transfer for qubit types currently integrated into quantum computers.

The results show that the power usage is extremely sensitive to values of $\phi$, which is consistent with previous findings that moving electronics inside the cryostat tended to exceed the cooling capacity of laboratory experiments. For a superconducting qubit, the energy used for cooling dominates over the electronics energy usage at values of $\phi$ ranging from $10^{-7}$ to $10^{-6}$, depending on the operating temperature. The cooling requirements of a system operating at 10 mK are roughly twice those of a system operating at 20 mK. For an ion-trap system, the cooling power requirements dominate the electronics power requirements at values of $\phi$ of around $10^{-3}$. There is also much less variation over the range of possible operating temperatures.

Figure 4 shows the range of possible PUEs for silicon and diamond qubits. Because silicon qubits have been demonstrated over temperatures from 0.1 to 10.0 K, the potential values for PUE range over three orders of magnitude. Depending on the operating temperature, the cooling power usage becomes dominant anywhere from a value of $\phi$ of $10^{-5}$ to $10^{-3}$. For diamond qubits, the range of potential power usage is much narrower. These results show that the type of qubit used, the operating temperature, and how the electronics are integrated all dramatically impact the power required to cool quantum systems.

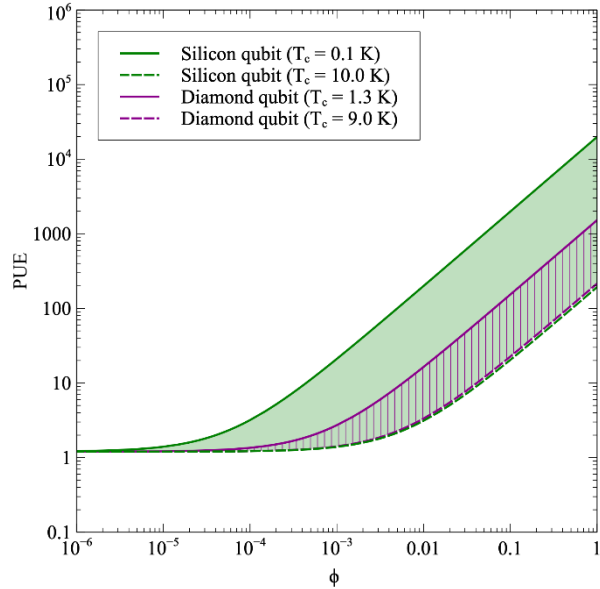

Fig. 4: Quantum Data Center Power Usage Efficiencies (PUEs) with no external heat transfer for qubit types currently demonstrated in laboratory environments.

### 4.1. Ideal Results: No heat transfer

For real systems, the heat transfer term $\beta$ will not be zero, meaning the PUE will no longer be independent of the system size, and $P^*$ will no longer be a linear function of the number of qubits. For any given value of $\phi$, $\beta$, or $n_p$, the shares of the power used (1) between the electronics $F_e$, (2) to cool the low-temperate electronics inside the cryostat $F_{lte}$, (3) to maintain the cryostat at temperature because of external heat transfer $F_o$, and (4) for external cooling $F_{ext}$ can be determined. Each of these can be written as a fraction of the overall power usage as found in Equation (16):

$$F_e \;=\; \frac{1}{\left[1 + \phi\left(\dfrac{1 + \beta n_p^{-1/3}}{\eta_c COP(T_c)|_C}\right) + \left(\dfrac{1 - \phi}{FOM(T_o)}\right)\right]}, \tag{18.a}$$

$$F_{lte} = \frac{[\phi/\eta_c COP(T_c)|_C]}{\left[1 + \phi\left(\frac{1+\beta n_p^{-1/3}}{\eta_c COP(T_c)|_C}\right) + \left(\frac{1-\phi}{FOM(T_o)}\right)\right]}, \tag{18.b}$$

$$F_o = \frac{[\phi\beta n_p^{-1/3}/\eta_c COP(T_c)|_C]}{\left[1 + \phi\left(\frac{1+\beta n_p^{-1/3}}{\eta_c COP(T_c)|_C}\right) + \left(\frac{1-\phi}{FOM(T_o)}\right)\right]}, \tag{18.c}$$

$$F_o = \frac{[(1-\phi)/FOM(T_o)]}{\left[1 + \phi\left(\frac{1+\beta n_p^{-1/3}}{\eta_c COP(T_c)|_C}\right) + \left(\frac{1-\phi}{FOM(T_o)}\right)\right]}. \tag{18.d}$$

Because of the complexity of the parameter space, the current results are limited to a value of $T_c$ of 15 mK, which represents a superconducting qubit system, and a value of $T_c$ of 4.5 K, which represents an ion-trap qubit system. For each case, values of $\phi$ of 0.1, 0.001, and 0.00001 with values of $\beta$ of 0, 0.01, 0.1, 1.0, 10.0, and 100.0 are considered for computers ranging in size from 10 to 1,000,000 qubits

The results for a value of $T_c$ of 15 mK are shown in Fig. 5. The resulting $P^*$ and PUE values show that the external heat transfer only changes the power usage when $\beta$ is greater than 1. They also show the effect of external heat transfer diminishes as the number of physical qubits increases. The magnitude of these effects appears to also depend on how the electronic power usage is split between external electronics and the cryostat electronics.

As shown in Fig. 5(c), for a system with a $n_p$ of 1,000 qubits and a $\phi$ of 0.1 operating at 15 mK, the power for cooling the electronics $F_{lte}$ inside of the cryostat is more than 99% of the total power usage for values of $\beta$ of up to 0.01. At a value of $\beta$ of 1, the share of power going to cool electronics drops to around 90%, and the share of power going to cooling for external heat transfer $F_o$ increases to around 10%. At a value of $\beta$ of 10, the two cooling needs are equal, while at a $\beta$ of 100, cooling for external heat transfer dominates. The combined power needs for all electronics and external cooling $F_e + F_{ext}$ never exceed 0.01% of the total power used. At a value of $\phi$ of 0.001, the power distributions are similar.

However, when $\phi$ is reduced to 0.0001, other forms of power usage become significant. When there is no external heat transfer, cooling the electronics in the cryostat requires approximately 50% of the total power usage, while the electronics themselves use 40% of the total power, and external cooling uses approximately 10% of the total power. However, as $\beta$ increases, the power required to remove heat from the cryostat because of external heat transfer dominates the power usage.

Figure 6 shows the same information for an ion-trap system operating at a temperature of 4.5 K. Because $COP_c(T_c)$ is approximately 300 times higher at 4.5 K than at 15 mK, the values for $P^*$ and PUE decrease significantly. The distribution of energy usage within the system also changes. At a value of $\phi$ of 0.1, the combined cooling loads for the electronics in the cryostat and external heat transfer into the cryostat are still dominant.

However, at a value of $\beta$ of 0, electronics account for 2.3% of total power use. This decreases to 1.1% of total power usage as $\beta$ increases to 10.0 and to less than 0.2% at a value of $\beta$ of 100.0, which reflects the decreased cost of cooling in the overall power budget.

As $\phi$ decreases to 0.001, the total cost of maintaining the low-temperature circuits at 4.5 K decreases dramatically. For values of $\beta$ of 10 or less, the electronics become the dominant consumer of power in the system. When $\phi$ decreases to 0.00001, the cost of maintaining low-temperature circuits at 4.5 K is only significant for values of $\beta$ greater than 10.0, and even then, it is less than 4% of total power usage. For these cases, $P^*$ and PUE do not change significantly as $\beta$ increases. The combination of reducing electronics in the cryostat to a minimal level and operating at higher temperatures lead to power usage that mimics a conventional computer. However, this configuration may compromise the ability of the conventional electronics to control the qubits effectively.

## 5. Quantum Volume Versus Power Scaling

In conventional computing, the power of a processor scales directly with the number of transistors, and the power of a parallel computer scales directly as the number of processors times some parallelization efficiency. For linear algebra problems, this can be quantified using the time required to solve standardized problems such as the commonly used LINPACK benchmark.[40] While there is no universally accepted equivalent for quantum computing, the concept of "quantum volume," which has been quantified for systems, has gained some acceptance.[41-42]

The quantum value $V_Q$ depends on the number of qubits $n_q$ and the effective error rate $\varepsilon_{eff}$. The effective error will depend on the computing architecture, the noise of the qubits, and the total number of qubits, and is therefore not completely independent of $n_q$. The quantum volume for a system of $n_q$ qubits with an effective error rate rate $\varepsilon_{eff}$ is given by:

$$V_Q = \max_{n<n_p}\left(min\left[n, \frac{1}{n\varepsilon_{eff}(n)}\right]^2\right). \tag{19}$$

This expression shows that the quantum volume is proportional to $n_p^{1/2}$ until the threshold value of $n_{peak}$ equal to $\sqrt{1/\varepsilon_{eff}(n)}$ is reached. At this point, adding additional qubits to the system will not increase the computational volume.

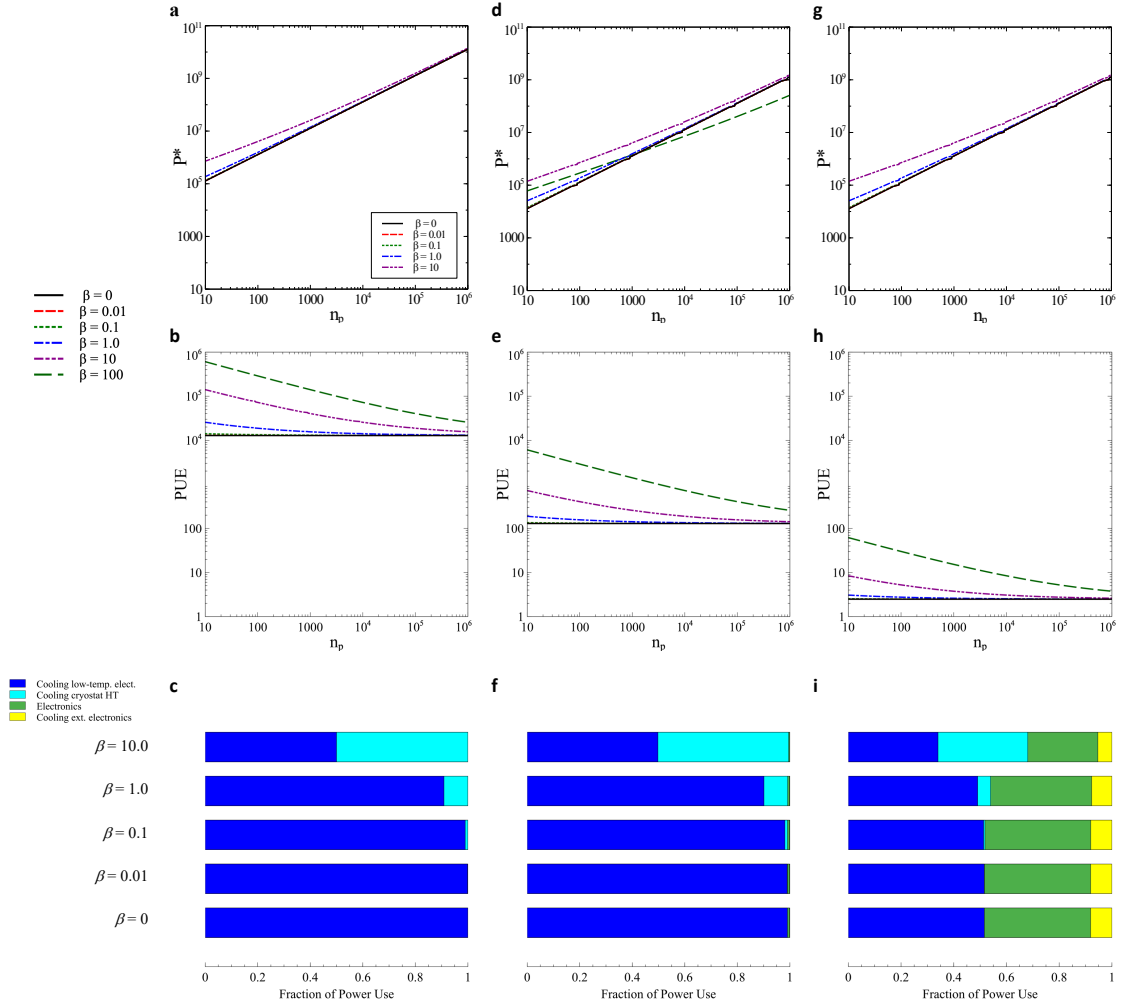

Fig. 5. Power usage, power usage efficiency (PUE), and power distribution for a quantum computer operating at a temperature of 15 mK. **a, b** $P^*$ and PUE for a system with $\phi$ of 0.1 and assorted values of $\beta$ as a function of $n_p$. **c** Power distribution for a system with $n_p$ of 1,000 and $\phi$ of 0.1. **d, e** $P^*$ and PUE for a system with $\phi$ of 0.001 and assorted values of $\beta$ as a function of $n_p$. **f** Power distribution for a system with $n_p$ of 1,000 and $\phi$ of 0.001. **g, h** $P^*$ and PUE for a system with $\phi$ of 0.00001 and assorted values of $\beta$ as a function of $n_p$. **i** Power distribution for a system with $n_p$ of 1,000 and $\phi$ of 0.00001.

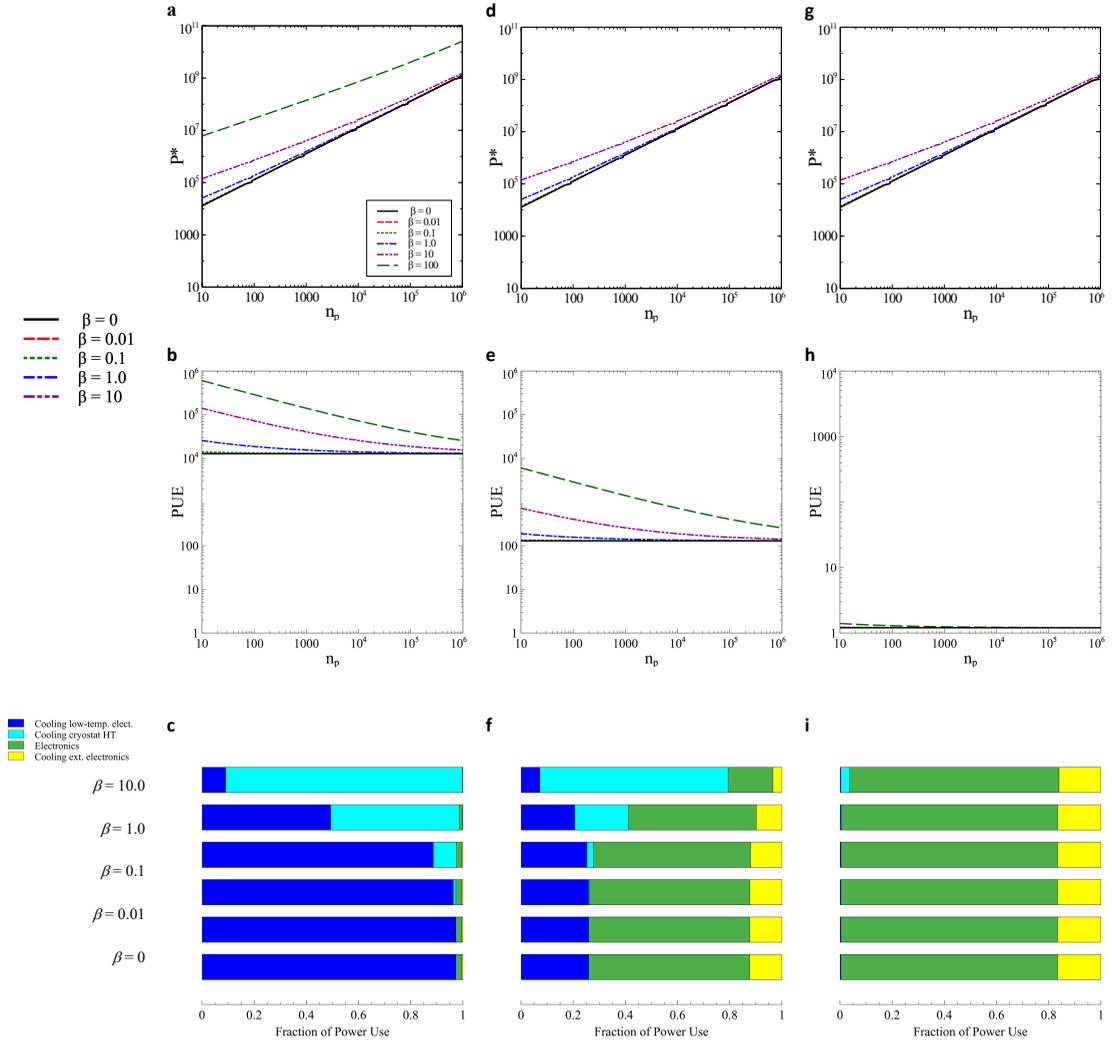

Fig. 6. Power usage, power usage efficiency (PUE), and power distribution for a quantum computer operating at a temperature of 4 K. **a, b** $P^*$ and PUE for a system with $\phi$ of 0.1 and assorted values of $\beta$ as a function of $n_p$. **c** Power distribution for a system with $n_p$ of 1,000 and $\phi$ of 0.1. **d, e** $P^*$ and PUE for a system with $\phi$ of 0.001 and assorted values of $\beta$ as a function of $n_p$. **f** Power distribution for a system with $n_p$ of 1,000 and $\phi$ of 0.001. **g, h** $P^*$ and PUE for a system with $\phi$ of 0.00001 and assorted values of $\beta$ as a function of $n_p$. **i** Power distribution for a system with $n_p$ of 1,000 and $\phi$ of 0.00001.

Figure 7 shows $V_Q$ as a function of the number of qubits for various error rates. These effective error rates are generally far below those seen in current machines: current two-qubit error rates for state-of-the-art superconducting processors with tens of physical qubits are in the range of 0.36%–1.1%,[3,42] while two-qubit error rates for trapped ion processors with a few to roughly a dozen physical qubits are in the range of 0.79%–2.5%.[43-44]

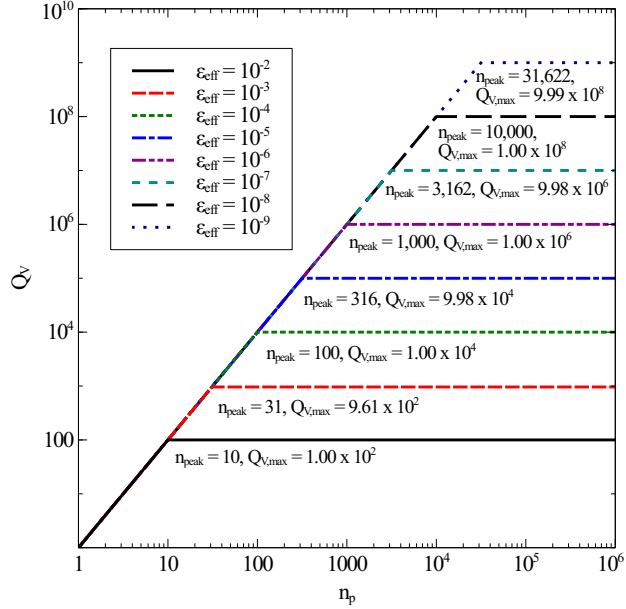

Fig. 7. Quantum volume as a function of qubits and effective noise

As long as the number of qubits is below $n_{peak}$, the quantum volume will be proportional to $n_p^2$, and equation (16) for the scaled power use can be re-written as:

$$P_T^* = Q_V^{1/2}\left[1 + \left(\frac{\phi}{\eta_c COP(T_c)|_C}\right) + \left(\frac{1-\phi}{FOM(T_o)}\right)\right] + Q_V^{1/6}\left[\frac{\phi\beta}{\eta_c COP(T_c)|_C}\right]. \quad (20)$$

As the limit of a large system, or negligible heat transfer is reached, the power required to operate a quantum computer scales with $V_Q^{1/2}$.

The scaled power use as a function of quantum volume for asystem with a $\phi$ of 0.001 operating at 0.015 K is shown as Fig. 8. The maximum values of $Q_V$ for different error rates are shown as vertical lines. The large system limit, where cryostat heat transfer can be ignored, is generally above the limit imposed by quantum volume consideration scalings.

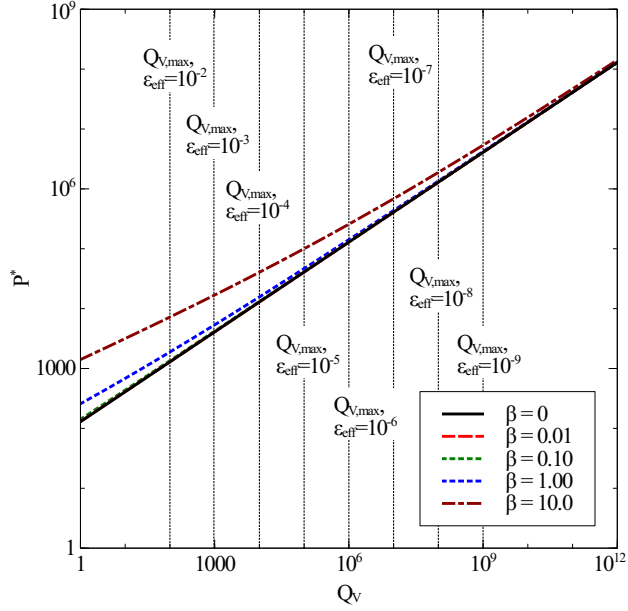

Fig. 8. Scaling of Power Use $P^*$ versus Quantum Volume $Q_V$ for a system withof 0.001 operating at 0.015 K. Vertical lines indicate the maximum achievable quantum volume for an effective error rate $\varepsilon_{eff}$

As shown in fig. 9, a similar pattern, with lower overall power usage, is obtained for a system operating at 4 K.

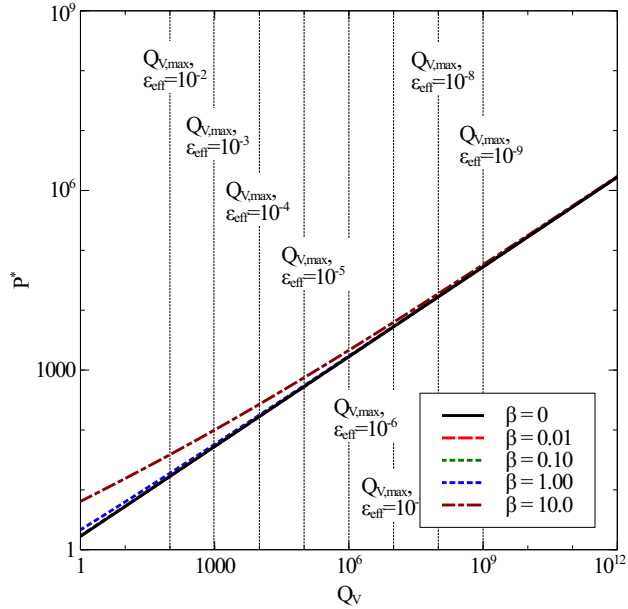

Fig. 9. Scaling of Power Use $P^*$ versus Quantum Volume $Q_V$ for a system withof 0.001 operating at 4.0 K. Vertical lines indicate the maximum achievable quantum volume for an effective error rate $\varepsilon_{eff}$

## 6. Conclusions

Quantum information systems use energy in very different ways than conventional computers. Instead of the energy use being dominated by energy used within the electronic circuitry, energy use is dominated by cooling requirements. While this is well understood empirically, the current work quantifies the requirements across computational architectures, computer sizes, and thermal design parameters. This work also points to two strategies for improving the energy efficiency of quantum systems, which will both lessen the energy impact of deploying quantum systems and reduce operational costs.

The first strategy is to reduce the energy costs of cooling to quantum temperatures. Within conventional data center operation, experience shows that careful engineering design and integration to reduce non-computing data center energy costs can reduce PUE from 1.60 to 1.02. This reduction corresponds to a thirtyfold reduction in noncomputational energy costs. The current work shows that the largest opportunity for reducing energy use and obtaining similar improvements in quantum systems is through improving $\eta_c$. This approach will require both careful consideration of cycle efficiency and efficient detailed design of individual components.

The second strategy is to design quantum systems in an energy-efficient manner. The scalings presented in this work show the importance of not only the computer size as represented by the number of qubits $n_p$ and the energy usage per qubit $q$, but also the operating temperature $T_c$ and the split of energy between quantum computing and conventional computing $\phi$. In addition to the inherent relative energy efficiencies gained by computers with large numbers of qubits, larger cryogenic systems appear to have higher efficiency values $\eta_c$. The number of physical qubits and the operating temperature are also expected to be factors in determining both the computing power and the amount of computing power obtained per unit energy usage. The physical packaging constraint $v_q$ determines cryostat size, and the impact of heat losses on system efficiency. Finally, the noise of the physical qubits and the error correction scheme used have such a large impact on computer scaling, that improving these is also a pathway to improved energy efficiency. As quantum technology matures, configurations can likely be identified that maximize computing power and minimize overall energy use.

The approach most likely to yield energy-efficient quantum systems is to combine these two strategies through *codesign*, where the impact of decisions made in deciding the computational architecture on the energy and cooling requirements of the system are considered during design and are matched with existing cooling capabilities. This may require designers to choose the optimal approaches to achieving a computational output within a set energy budget by selecting system architecture parameters.

**Acknowledgment**

This work was authored by the National Renewable Energy Laboratory (NREL), operated by Alliance for Sustainable Energy, LLC, for the U.S. Department of Energy (DOE) under Contract No. DE-AC36-08GO28308. This work was supported by the Laboratory Directed Research and Development (LDRD) Program at NREL. The views expressed in the article do not necessarily represent the views of the DOE or the U.S.